\documentclass[submission, Phys]{SciPost}

\newcommand{\be}{\begin{equation}}
\newcommand{\ee}{\end{equation}}
\newcommand{\bea}{\begin{eqnarray}}
\newcommand{\eea}{\end{eqnarray}}
\usepackage{braket}
\usepackage{amsfonts}
\usepackage{soul}
\usepackage{cleveref}
\newcommand{\w}{\omega}
\newcommand{\Id}[1] {\int_{-\infty}^{\infty} \! \! {\rm d} #1}

\begin{document}

\begin{center}{\Large \textbf{
Spectral and transmission properties of multiple correlated quantum dots made simple
}}\end{center}

\begin{center}
N. Sobrino\textsuperscript{1*},
S. Kurth\textsuperscript{2,3,4}
\end{center}

\begin{center}
  {\bf 1}     The Abdus Salam International Centre for Theoretical Physics
  (ICTP), Strada Costiera 11, 34151 Trieste, Italy
\\
  {\bf 2} Nano-Bio Spectroscopy Group and European Theoretical Spectroscopy
  Facility (ETSF), Departamento de Pol\'imeros y Materiales Avanzados:
  F\'isica, Qu\'imica y Tecnolog\'ia, Universidad del Pa\'is Vasco UPV/EHU,
  Avenida de Tolosa 72, E-20018 San Sebasti\'an, Spain
\\
  {\bf 3} IKERBASQUE, Basque Foundation for Science, Plaza Euskadi 5,
  E-48009 Bilbao, Spain
\\
  {\bf 4} Donostia International Physics Center (DIPC), Paseo Manuel de
  Lardizabal 4, E-20018 San Sebasti\'{a}n, Spain
\\
* nsobrino@ictp.it
\end{center}

\begin{center}
\today
\end{center}


\section*{Abstract}
{\bf
  Steady-state density functional theory, called i-DFT, is employed to compute
  spectral and transmission properties of general interacting nanoscale
  regions coupled to electronic reservoirs. Exchange-correlation functionals
  are constructed for different interactions and coupling geometries.
  The potential of the method is illustrated by applications to various
  multiple quantum dots from the Coulomb blockade to the Kondo regime.
  The results are in reasonable agreement with many-body approaches at a fraction of the
  computational cost.

}

\vspace{10pt}
\noindent\rule{\textwidth}{1pt}
\tableofcontents\thispagestyle{fancy}
\noindent\rule{\textwidth}{1pt}
\vspace{10pt}

\section{Introduction}
\label{sec:intro}



Quantum dots connected to metallic leads provide a unique
testbed for both experimental and theoretical studies of strong electronic 
correlations \cite{Wiel_etal:02,kouwenhoven1997electron}. On the theoretical side, such systems 
are typically modelled in terms of simple tight-binding Hamiltonians
and their spectral and transport properties can be studied with a range of
techniques with different merits and drawbacks. Rate equations
\cite{Beenakker:91} as well as simple truncations of the equations of motion (EOM) of the Green function 
\cite{hubbard1964electronII,sierra2016interactions,SobrinoJacobKurth:24,SobrinoJacobKurth:25,Sobrino:25,chang2008theory} in the Coulomb blockade regime already provide the correct
physical picture at small numerical cost. However, for going beyond this
regime other truncation schemes or numerically more demanding techniques have to be employed such
as, e.g., numerical and functional renormalization
group (NRG and fRG), density matrix renormalization group (DMRG), Quantum
Monte Carlo (QMC), many-body perturbation theory (MBPT), hierarchical equations of motion (HEOM) as well as
expansions around the atomic limit (non-crossing and one-crossing
approximations, NCA and OCA) \cite{bulla2008numerical,izumidasakai:05,karrasch2006functional,EckelHeidrichJakobsThorwartPletyokhovEgger:10,schollwock2011density,gull2011continuous,haug2008quantum,bickers1987review,pruschke1989anderson,melo2020impuritysolvers,tanimura2020numerically}.

The simplest model of a QD connected to metallic leads is the Single Impurity
Anderson Model (SIAM). Spectral and transport properties of this model (both
in and out of the Kondo regime) have been calculated with all of the methods
mentioned above, among others. 

While studies of the SIAM abound, double or multiple QDs have been studied less frequently, specially in the Kondo
regime. For double quantum dots (DQDs), several studies have addressed strongly correlated regimes, including works based on NRG 
\cite{IzumidaSakaiShimizu:98,BoeseHofstetterSchoeller:01,keller2014emergent,Wang:11,Weymann_etal:18,kleeorin2018quantum,Galpin2005,Galpin2006,Zitko2007,Zitko2006}
  and NCA \cite{LombardoHaynZhuravelSchaefer:20,DarocaRouraAligia:25,daroca2025thermoelectric,perez2023thermoelectric}.
Already at the DQD level, the physics is qualitatively richer than in the SIAM: interdot tunnelling, capacitive coupling, interdot and exchange interactions give rise to molecular bonding/antibonding states, two-impurity and SU(4)  Kondo correlations, RKKY-mediated exchange, and quantum phase transitions between distinct strongly correlated ground states~\cite{Wiel_etal:02,Weymann_etal:18,mravlje2006kondo,chung2007kondo,Wang:11,kleeorin2018quantum}, among other phenomena. These studies have also made  clear that capturing the relevant physics requires access not only to local spectral weights but also to the off-diagonal elements of the many-body spectral function matrix, which encode interference between dots and ultimately determine the measured transmission. Multiple quantum dots (MQDs) with three or more dots bring additional phenomena into play such as charge frustration in non-trivial geometries, higher SU($N$) Kondo correlations, or more involved quantum-phase-transition landscapes ~\cite{seo2013charge,zitko2006kondo,numata2009kondo,lopez20133,mitchell2010two}. These MQD systems have been explored considerably less, since the available methods are constrained by their underlying approximations and by a numerical cost that grows rapidly with system size.
  
This motivates the search for a complementary framework whose numerical cost remains manageable as the number of dots grows, while still giving access to the full many-body transmission and spectral functions including in the strongly correlated regime.

In the present work, we use a particular flavor of density functional theory (DFT) for the computation of spectral 
properties of MQDs. While DFT is usually not associated with a proper description of strong correlations, it has 
been realized some time ago that the Kondo effect in the SIAM can be captured within a standard DFT description, provided a 
functional with a step feature at half filling is used 
\cite{StefanucciKurth:11,BergfieldLiuBurkeStafford:12,TroesterSchmitteckertEvers:12}. 
Later, a general DFT framework for the description of electronic transport, named steady-state DFT (or i-DFT), has
been established \cite{StefanucciKurth:15} which allows for a proper description of the transport characteristics 
of the SIAM both in the linear and non-linear bias regimes \cite{KurthStefanucci:16}. In a particular 
limiting case (called the {\em ideal STM limit}), i-DFT can also be used
to compute spectral functions in thermal equilibrium both for QDs
\cite{JacobKurth:18,KurthJacobSobrinoStefanucci:19} and for bulk systems\cite{JacobStefanucciKurth:20,JacobKurth:26}. A related extension, termed iq-DFT, also has
been formulated to include a proper description of the heat current by enlarging the basic set of variables from the density and charge current to the 
density, charge current, and heat current \cite{SobrinoEichStefanucciDAgostaKurth:21,sobrino2023thermoelectric}.

In the present paper, we exploit the ideal STM limit to access spectral
properties of MQDs. We generalize the work of Ref.~\cite{JacobKurth:18} 
in various directions: (i) we construct approximations for 
the (H)xc gate and bias potentials for previously not accessible MQDs with 
general local and intersite density-density interactions, (ii) 
we extend the framework to off-diagonal coupling matrices which allows to extract 
off-diagonal spectral information which is needed to compute, e.g., transmission 
spectral functions and allows for an {\em analytical} understanding of 
intereference-related antiresonances in this quantity, and (iii) we formulate  
the KS problem in terms of the biorthogonal eigensystem of a non-Hermitian 
Hamiltonian which allows to evaluate all frequency integrals in the equations for 
density and current analytically, leading to a highly efficient implementation 
of the KS self-consistent cycle. The structure of the paper is as
follows: in the next Section, we spell out the model Hamiltonian for the
MQDs we will study and give a brief introduction to i-DFT and the ideal
STM limit. We derive an equation which expresses many-body spectral
properties completely in terms of i-DFT quantities for arbitrary coupling to
the leads. In Sec.~\ref{sec:applications} we show applications to both local
spectral functions as well as transmission spectral functions of MQDs. We
explain how we construct approximate exchange-correlation bias functionals 
for different couplings and different interactions. In
Sec.~\ref{ssec:transmission_spectra} we investigate transmission spectral
functions of a double QD and analyze the
interference antiresonance which develops across the quantum phase transition reported for this system. In
Sec.~\ref{sec:conclusions} we summarize and present our conclusions.

\section{Model and Methodology}
\label{sec:model}

\subsection{Multiple Quantum Dots: Spectral Properties from Transport}
\label{ssec:mqd}

 We aim to describe general spectral properties of a multiple quantum dot
 (MQD) of interacting electrons in contact with non-interacting electronic reservoirs at temperature $T$. 
 The system is described by the Hamiltonian
 \be
 \hat{H}_0 = \hat{H}_{\rm MQD} + \hat{H}_{\rm res} + \hat{H}_{\rm C} 
 \label{hamil}
 \ee
 with the Hamiltonian of the MQD consisting of $M$ dots given as 
 \be
 \hat{H}_{\rm MQD}
 =
 \sum_{l=1}^M v_l \hat{n}_l
 +
 \sum_{\stackrel{l,m}{l < m}}^M
 \sum_{\sigma= \uparrow,\downarrow}
 t_{lm}
 \left(
 \hat{d}^{\dagger}_{l\sigma}\hat{d}_{m \sigma}
 +
 H.c.
 \right)
 + \frac{1}{2}
 	\sum_{l,l',m,m'=1}^M
 	\sum_{\sigma,\sigma'}
 	W_{ll'mm'}\,
 	\hat{d}^{\dagger}_{l\sigma}
 	\hat{d}^{\dagger}_{l'\sigma'}
 	\hat{d}_{m'\sigma'}
 	\hat{d}_{m\sigma}
 \;.
 \label{hamil_dot}
 \ee

 Here the $\hat{d}^{\dagger}_{l \sigma}$ ($\hat{d}_{l \sigma}$) are the creation
 (annihilation) operators for an electron with spin $\sigma$ at site $l$,
 $\hat{n}_{l \sigma} =\hat{d}^{\dagger}_{l \sigma}\hat{d}_{l \sigma}$ is the
 spin density at site $l$, and
 $\hat{n}_l = \sum_{\sigma=\uparrow, \downarrow} \hat{n}_{l \sigma}$. 
 The $v_l$ are the on-site energies at site $l$, while the $t_{lm}$ are the hopping
 matrix elements between sites $l$ and $m$ and the $W_{ll'mm'}$ describe the two-body electron-electron interaction within the MQD. 
 Hamiltonians of this form describe, e.g., gate-defined
semiconductor quantum dots, carbon-nanotube and nanowire quantum dots, serial or
parallel double-dot devices, and molecular junctions in which a few localized
orbitals dominate low-energy transport. In gate-defined quantum-dot devices, charging energies, interdot interactions
and tunnel couplings can be extracted from charge-stability and Coulomb-blockade
measurements~\cite{kouwenhoven1997electron,Wiel_etal:02}, while for molecular
junctions the model parameters can be obtained from first-principles
calculations, with the interaction parameters computed, e.g., by constrained
RPA~\cite{aryasetiawan2004frequency} or linear response~\cite{cococcioni2005linear}.
 The methodology described below in principle holds for any interaction $W_{jklm}$, 
 but different forms in general require different approximations. Therefore, in the applications we will specialize 
 to on-site interactions as well as density-density interactions between sites. The reservoir of non-interacting
 electrons at vanishing chemical potential is described by
\be
\hat{H}_{\rm res} = \sum_{k, \sigma} \varepsilon_k \hat{c}^{\dagger}_{k \sigma}
\hat{c}_{k \sigma}
\ee
with the corresponding creation (annihilation) operators $\hat{c}_{k \sigma}$
($\hat{c}^{\dagger}_{k \sigma}$), while the coupling between the reservoir and the 
MQD is
\be
\hat{H}_{\rm C} = \sum_{l \sigma} \sum_k T_{lk} (\hat{d}^{\dagger}_{l \sigma} 
\hat{c}_{k \sigma} + H.c.) \; .
\ee
The tunnelling amplitudes $T_{lk}$ should not be confused with the scalar $T$, which throughout this work denotes the reservoir temperature. Since the reservoir is non-interacting, it may conveniently be treated with 
the standard embedding technique through the embedding self energy 
${\bf \Sigma}_{\rm res}(\w)$ which is a matrix in the single-particle basis 
spanning the Hilbert space of $\hat{H}_{\rm MQD}$. The broadening matrix 
corresponding to the embedding self energy is 
${\bf \Gamma}(\w) = i ({\bf \Sigma}_{\rm res}(\w) - 
{\bf \Sigma}_{\rm res}^{\dagger}(\w))$ with matrix elements 
${\bf \Gamma}_{lm}(\w) = 2 \pi \sum_k T_{lk} T_{mk}^* \delta(\w-\varepsilon_k)$. 
We work in the wide band limit and assume these matrix elements to be 
independent of frequency, i.e., ${\bf \Gamma}_{lm}(\w) = \gamma_{lm}$. 

In order to access the spectral properties of the MQD in contact with the
reservoir we use a scheme proposed in Ref.~\cite{JacobKurth:18}. The idea
is to connect as a probe $P$ a second (wide band) lead with broadening
matrix ${\mathbf \Gamma}_{\rm P}$ of non-interacting
electrons to the system of interest and bias this second lead with bias $V$ to
drive a (steady) current $I$ through the system. Using the Meir-Wingreen
formula \cite{MeirWingreen:92}, one can then show that, if the probe is at zero
temperature and the coupling of the probe is infinitesimally weak, the
resulting differential conductance is related to the equilibrium spectral
function matrix ${\mathbf A}$ of the interacting system via
\be
\frac{{\rm d} I}{{\rm d} V} \xrightarrow[{\mathbf \Gamma}_{\rm P}\to0]{}
\frac{{\rm Tr}\{ {\mathbf \Gamma}_{\rm P} {\mathbf A}(V)\}}{\pi} 
\label{dIdV_STM}
\ee
where the trace is to be taken over a complete set of single-particle basis
functions of the interacting region. The setup with an infinitesimally weakly
coupled probe lead has been termed the ``ideal STM limit''
\cite{JacobKurth:18} since the probe resembles the tip of a scanning tunneling
microscope. Two points are important to highlight here: (i) the structure of
the coupling matrix ${\mathbf \Gamma}_{\rm P}$ is flexible and can be chosen
according to the quantity one aims to compute and (ii) one may use
Eq.~(\ref{dIdV_STM}) to extract equilibrium spectral features 
of the system of interest with {\em any} theoretical framework which allows to
compute the steady current through a biased interacting system. 
One such framework, steady-state density functional theory (or i-DFT), will
be discussed next.

\subsection{Steady-State Density Functional Theory (i-DFT)}
\label{ssec:idft}

Density functional theory (DFT) is an ubiquitous method for the {\em ab initio}
description of atoms, molecules and solids. In its original incarnation
\cite{HohenbergKohn:64,KohnSham:65}, it describes the ground state or thermal
equilibrium state \cite{Mermin:65} of an interacting many-electron system
solely in terms of its (ground state or equilibrium) electronic density
\cite{DreizlerGross:90}. A DFT framework to describe systems driven out of
thermal equilibrium (such as transport) is time-dependent DFT
\cite{RungeGross:84,Ullrich:12} where one follows the dynamical evolution
in response to a time-dependent potential (e.g., a time-dependent bias).
Here we use an alternative approach specifically designed to describe
steady-state transport, namely steady-state DFT or i-DFT
\cite{StefanucciKurth:15}. The central idea of i-DFT is to describe a
many-electron system connected to two biased leads in its steady state in
terms of both the (steady state) density on and the steady current through the
device. Unlike standard DFT, the auxiliary non-interacting Kohn-Sham (KS)
system of i-DFT involves not only an Hartree-exchange-correlation (Hxc)
contribution to the effective potential in the device but also an
exchange-correlation (xc) contribution to the bias. These (H)xc ``potentials''
are functionals of both the density and the current. Just like standard DFT,
i-DFT has been formulated for an {\em ab initio} description of electron
transport. However, so far i-DFT has successfully been
applied to lattice models with particular emphasis on effects of strong
electronic correlations \cite{KurthStefanucci:17}. In particular,
the single-impurity Anderson model (SIAM) has been studied from the Coulomb
blockade \cite{StefanucciKurth:15} to the Kondo regime
\cite{KurthStefanucci:16,SobrinoAgostaKurth:19}. 

We consider a lattice model with a central interacting region of $M$ sites
connected to left ($L$) and right ($R$) leads with a total bias $V$
symmetrically applied across the central region and temperatures $T_{\alpha}$
for lead $\alpha$. Then the KS equations for the site densities (occupations)
$n_l$ of and the steady
current $I$ through the central region are
\be
n_l = \frac{1}{\pi} \sum_{\alpha \in \{{\rm L,R}\}} \Id{\w} \; f_{\alpha}(\w)
\left[ {\mathbf G}^{R}(\w) {\mathbf \Gamma}_{\alpha}(\w) {\mathbf G}^{A}(\w)
\right]_{ll} \mbox{\hspace*{1cm} for $l\in \{ 1, \ldots,M\}$}
\label{densities}
\ee
and
\be
I = \frac{1}{\pi} \Id{\w} \; \left[ f_{\rm L}(\w) - f_{\rm R}(\w) \right]
{\rm Tr}\{ {\mathbf \Gamma}_{\rm L}(\w) {\mathbf G}^{R}(\w)
{\mathbf \Gamma}_{\rm R}(\w) {\mathbf G}^{A}(\w)
\} \;. 
\label{current}
\ee
Here, ${\mathbf G}^{R}(\w)$ (${\mathbf G}^{A}(\w)={\mathbf G}^{R,\dagger}(\w)$)
are the retarded (advanced) KS Green functions of the central region in the
site basis given by
\be
{\mathbf G}^{R}(\w) = \left[ \w {\mathbf 1} - {\mathbf H} -
  {\mathbf \Sigma}(\w) \right]^{-1}
\label{ks_gf}
\ee
with the $M \times M$ unit matrix ${\mathbf 1}$, and the KS Hamiltonian matrix
${\mathbf H}$ whose diagonal elements are given by
\be
{\mathbf H}_{ll} = v_{s,l} = v_l + v_{\rm Hxc,l}(\{n_m\},I) \;.
\ee
Here, $v_l$ is the on-site energy at site $l$ and the $v_{\rm Hxc, l}$ is the
corresponing Hxc potential which depends on the occupations $n_m$ at all
sites as well as on the current $I$.  The off-diagonal elements of the
KS Hamiltonian are ${\mathbf H}_{lm} = t_{lm}$, see Eq.~(\ref{hamil_dot}). 
Furthermore, we have defined the total embedding self energy
${\mathbf \Sigma}(\w)={\mathbf \Sigma}_{\rm L}(\w-V_{s}/2) +
{\mathbf \Sigma}_{\rm R}(\w+V_{s}/2)$
with the corresponding embedding self energies ${\mathbf \Sigma}_{\alpha}$
and broadening matrices
${\mathbf \Gamma}_{\alpha}(\w) =i ( {\mathbf \Sigma}_{\alpha}(\w) - 
{\mathbf \Sigma}^{\dagger}_{\alpha}(\w) )$ of lead $\alpha$. Finally,
$f_{{\rm L}/{\rm R}}(\w) = [ 1 + \exp((\w \mp V_{s}/2)/T_{{\rm L}/{\rm R}} ]^{-1}$
is the Fermi function for lead $\alpha$ with the KS bias
\be
V_{s} = V + V_{\rm xc}(\{n_m\},I)
\label{xcbias}
\ee
consisting of the external part and the xc contribution to the bias,
$V_{\rm xc}$, which just as the Hxc potential $v_{\rm Hxc, l}$ depends on all
occupations as well as on the current. Note that due to the dependence
of both $v_{\rm Hxc, l}$ and $V_{\rm xc}$ on {\em all} occupations $n_m$ as well
as on the steady current $I$, Eqs.~(\ref{densities}) and (\ref{current}) are
coupled and have to be solved together.

\subsection{i-DFT in the STM limit}
\label{ssec:idft_stm}

As we have seen in Sec.~\ref{ssec:mqd}, in order to access spectral properties
of MQDs one needs to go to the ideal STM limit where the bias drops completely
at one of the leads, the infinitesimally weakly coupled probe (or tip). Within
i-DFT, this limit was first suggested and investigated in
Ref.~\cite{JacobKurth:18} while the continuous transition from symmetrically
to asymmetrically biased junctions was addressed in Ref.~\cite{KurthJacob:18}. 

In the following we take both leads to be in the wide band limit with the
left lead as the probe and the right lead taking the role as reservoir
(see Sec.~\ref{ssec:mqd}). We write the broadening matrix of the probe as
${\mathbf \Gamma_{\rm L}(\w)} = {\mathbf \Gamma_{\rm P}} = \gamma_{\rm P}
\tilde{\mathbf \Gamma}_{\rm P}$ with the real parameter $\gamma_{\rm P}$ and
the non-vanishing, dimensionless matrix $\tilde{\mathbf \Gamma}_{\rm P}$
independent of $\gamma_{\rm P}$. The right lead is the reservoir also taken
in the wide band limit, i.e.,
${\mathbf \Gamma_{\rm {R}}(\w)} = {\mathbf \Gamma}$.
Note that since both leads are taken as wide band, the total embedding self
energy then also becomes independent of frequency and purely imaginary, i.e.,
${\mathbf \Sigma(\w)} = -\frac{i}{2}({\mathbf \Gamma_{\rm P}}+{\mathbf \Gamma})$.
In the ideal STM limit ($\gamma_{\rm P} \to 0$), the term in the i-DFT equations
(\ref{densities}) related to the probe drops out and the equation becomes
identical to the DFT equations in equilibrium \cite{JacobKurth:18}, i.e.,
\be
n_l = \frac{1}{\pi} \Id{\w} \; f(\w)
\left[ {\mathbf G}^{R}(\w) {\mathbf \Gamma}(\w) {\mathbf G}^{A}(\w)
\right]_{ll} = \frac{1}{\pi} \Id{\w} \; f(\w) {\mathbf A}_s(\w)_{ll}
\mbox{\hspace*{5mm} for $l\in \{ 1, \ldots,M\}$}
\label{densities_stm}
\ee
where we used
\be
{\mathbf G}^{R}(\w) {\mathbf \Gamma}(\w) {\mathbf G}^{A}(\w) =
{\mathbf A}_s(\w)
\ee
with the KS spectral function matrix is 
${\mathbf A}_s(\w) = i ({\mathbf G}^{R}(\w) - {\mathbf G}^{A}(\w) )$ 
and $f(\w) = [ 1 + \exp(\w/T_{{\rm R}}) ]^{-1}$ is the Fermi
distribution of the reservoir. Note that the i-DFT Hxc potential entering
the KS Green functions in the STM limit reduces to standard DFT Hxc potential
at equilibrium \cite{JacobKurth:18}. The equilibrium KS Green function entering
Eq.~(\ref{densities_stm}) can be written as
\begin{equation}
	{\mathbf G}^{R}(\omega)		=
		\left[\omega{\mathbf 1}-{\mathbf H}_{\rm eff}\right]^{-1}, \qquad
        {\mathbf H}_{\rm eff} =
		{\mathbf H}_s-\frac{i}{2}{\mathbf \Gamma}.
	\label{eq:Gs_stm}
\end{equation}
Since \({\mathbf H}_{\rm eff}\) is in general non-Hermitian, we use its
bi-orthogonal spectral representation (as in the wide-band-limit NEGF treatment
of non-interacting junctions~\cite{tuovinen2014time,ridley2022many}). Denoting by \(w_k\) its eigenvalues,
with right and left eigenvectors \(\mathbf{r}_k\) and \(\mathbf{l}_k\) satisfying
	\[
	{\mathbf H}_{\rm eff}\,\mathbf{r}_k=w_k\,\mathbf{r}_k,\qquad
	{\mathbf H}_{\rm eff}^\dagger\,\mathbf{l}_k=w_k^*\,\mathbf{l}_k,
	\]
for a diagonalizable \({\mathbf H}_{\rm eff}\), the resolvent admits the 
spectral representation (see Appendix~\ref{sec:appendix1} for the derivation)
\begin{equation}
	({\mathbf G}^{R}(\omega))_{lm}	=
	\sum_k\frac{r_{k,lm}}{\omega-w_k}, \qquad r_{k,lm} =
	\frac{(\mathbf{r}_k)_l\,(\mathbf{l}_k)_m^*}
	{\mathbf{l}_k^\dagger\,\mathbf{r}_k}.
	\label{eq:poles_res_stm}
\end{equation}
Substituting Eq.~(\ref{eq:poles_res_stm}) into	Eq.~(\ref{densities_stm}) gives the closed form expressions for the densities
\begin{equation}
	n_l = 1-\frac{2}{\pi}
	\Phi(\mu=0,T_{\rm R};\{r_{k,ll}\},\{w_k\}).
	\label{eq:n_digamma_stm}
\end{equation}
with the auxiliary function
\begin{equation}
	\Phi(V,T;\{r_k\},\{w_k\})	=
	\sum_k {\rm Im}\!\left[ r_k\,
	\psi\!\left( \frac{1}{2}+\frac{i}{2\pi T}(w_k-V) \right)
	\right],
	\label{eq:Phi_general_stm}
\end{equation}
where \(\psi\) is the digamma function.

On the other hand, the i-DFT equation for the current in the STM limit becomes
\bea
I 
&=& \lim_{{\mathbf \Gamma}_{\rm P} \to 0}
\frac{1}{\pi} \Id{\w} \; \left[ f_{\rm P}(\w) - f(\w) \right]\tau_s(\w)
\label{current_stm}
\eea
with $\tau_s(\omega)={\rm Tr}\{{\mathbf \Gamma}_{\rm P}{\mathbf A}_s(\omega)\}$, and
the Fermi function of the probe (with $T_{\rm P}=0$) 
$f_{\rm P}(\w) = [ 1 + \exp((\w-V_s)/T_{{\rm P}}) ]^{-1}$. Of course, in the STM limit
($\gamma_{\rm P} \to 0$) the current goes to zero but $I/\gamma_{\rm P}$ remains
finite.

From Eq.~(\ref{eq:poles_res_stm}) one has
	\[
	{\rm Tr}\{
	{\mathbf \Gamma}_{\rm P}{\mathbf G}^{R}(\omega)
	\}
	=
	\sum_k\frac{q_k}{\omega-w_k},
	\qquad
	q_k
	=
	\sum_{ij}
	({\mathbf \Gamma}_{\rm P})_{ji}r_{k,lm}.
	\]
Using the same function \(\Phi\) defined in Eq.~(\ref{eq:Phi_general_stm}), the current can therefore be written as
\begin{equation}
	I 	= \frac{2}{\pi}
	\left[ \Phi(V=0,T_{\rm R};\{q_k\},\{w_k\})
	- \Phi(V_s,T_{\rm P};\{q_k\},\{w_k\})
	\right].
\label{eq:Ired_digamma_stm}
\end{equation}

It is important to emphasize that in the STM limit, the equation for the current 
(Eq.~(\ref{current_stm}) or, alternatively, Eq.~(\ref{eq:Ired_digamma_stm})) is completely decoupled from the 
equations for the densities (Eqs.~(\ref{densities_stm}) or (\ref{eq:n_digamma_stm})). This means that one can 
first find the self-consistent solutions for the (equilibrium) densities $n_l$, and then, using the 
corresponding equilibrium Hxc potential $v_{\rm Hxc, l}$) evaluated at these densities, one solves the 
self-consistent equation for the current.

In Eq.~(\ref{dIdV_STM}) we have seen that in the STM limit the differential
conductance is related to spectral properties. We now express the interacting quantity
\be
\tau(\w) = {\rm Tr}\{ {\mathbf \Gamma}_{\rm P} {\mathbf A}(\w)\}
\label{tau_def}
\ee
in terms of i-DFT quantities by taking the derivative of
Eq.~(\ref{current_stm}). When calculating this derivative one has to take
into account that (i) the only dependence of the r.h.s. of
Eq.~(\ref{current_stm}) on the bias occurs through the dependence of the
xc bias $V_{\rm xc}$ (which enters as argument in $f_{\rm P}(\w)$) on the current
and (ii) since the probe is at temperature $T_{\rm P}=0$, the derivative of
$f_{\rm P}(\w)$ with respect to its argument becomes a negative Dirac delta
function. The calculation then gives the compact result 
\be
\tau(\w) = \frac{\tau_s(\w + V_{\rm xc})}{1 - \frac{1}{\pi}
	\frac{\partial V_{\rm xc}}{\partial I} \, \tau_s(\w + V_{\rm xc})}
\label{tau_idft}
\ee
where $V_{\rm xc}$ and its derivative are to be evaluated at the current $I(\w)$ corresponding to 
the external bias $V=\w$.

The STM limit of i-DFT  has so far been used mainly to obtain local
and total spectral functions of nanoscale systems such as the single impurity Anderson model (SIAM) or the
Constant Interaction Model (CIM) \cite{JacobKurth:18}, as well as spectral functions of extended systems such 
as the Hubbard model in more than one dimension \cite{JacobStefanucciKurth:20,JacobKurth:26}. The formulation presented here
provides a flexible and also more general route to access spectral information beyond purely local spectra 
since with different choices of couplings to reservoir one can also project out nonlocal spectral information.
As a consequence, the presented framework gives access to both local and transmission
spectral functions (for an exact definition of the latter quantity see Sec.\ref{ssec:transmission_spectra}) of arbitrary 
interacting MQD, as will be shown in Sec.~\ref{sec:applications}.

We would like to emphasize the computational efficiency of the present approach to compute spectral properties of MQDs with $M$ interacting sites. To obtain the 
densities, a self-consistent, DFT calculation is required which involves the solution of $M$ coupled equations, each 
iteration requiring only the diagonalization of the $M\times M$ matrix $\mathbf{H}_{\rm eff}$, scaling as $\mathcal{O}(M^3)$, and no numerical frequency 
integration. Once the densities are known, each spectral property is obtained at each frequency as the solution of one additional (and independent) 
nonlinear equation, and the memory requirement remains polynomial in $M$. In contrast, many-body approaches such as exact diagonalization operate in 
Fock space whose dimension grows exponentially with $M$ (as $4^M$), severely limiting the system sizes which can be treated. 
In NRG, each Wilson shell enlarges 
the Hilbert space by a factor $4^M$ for $M$ screening channels. For the NRG benchmarks of Sec.~3.2 this amounts to an increase of computing time by several orders of magnitude.. 
Of course, the caveat of our i-DFT method (just as for DFT in general) is the need for accurate approximations for the (H)xc functionals. However, in i-DFT it is 
relatively easy to design functionals which capture Kondo physics accurately where many-body reference results are available \cite{JacobKurth:18} and at least 
qualitatively where these are lacking. Moreover, the fact that the i-DFT expression (\ref{tau_idft}) for spectral properties is written in terms of effectively non-
interacting KS quantities also allows for analytical investigations and arguments, see, e.g., Sec.~\ref{ssec:transmission_spectra}.

\section{Applications}
\label{sec:applications}

\subsection{Exchange-correlation functionals}
\label{ssec:xcbias}

The framework presented in Sec.~\ref{ssec:idft_stm} is formally exact and through Eq.~(\ref{tau_idft}) one can compute spectral properties of MQDs. In practice, the quality of the results depends on the accuracy of the approximate functionals employed. We now present the approximations which we employed to obtain our results. 
In the following we restrict ourselves to interactions of density-density form, 
i.e., local onsite Hubbard repulsions $U_l=W_{llll}$ at site $l$ plus density-density interactions $U_{lm}=W_{lmlm}$ between sites $l$ and $m$. All other two-electron integrals are taken to vanish. For interactions of this form, Hxc potentials for MQDs not attached to leads have been derived in Ref.~\cite{SobrinoKurthJacob:20} where the MQDs are taken to be in contact with a heat and particle bath treated via a grand-canonical 
ensemble (GCE). At low temperature of the bath, the Hxc potentials were found to exhibit thermally broadened step structures. Here we use Hxc potentials with the same step structures but now the broadening of the steps is due to the coupling to the leads. 
To be explicit, we work in what in Ref.~\cite{SobrinoKurthJacob:20} has been defined as parameter Regime I for which any on-site interaction is larger or equal than any intersite interaction, i.e., $U_k \geq U_{lm}$ for any $k$ and $l\neq m$. In this regime, the Hxc potential for site $l$ can be written as a sum of a SIAM and a CIM contribution~\cite{SobrinoKurthJacob:20}:
\be
v_{\rm Hxc,l}(\{n_m\}) = v^{\rm CIM}_{\rm Hxc}(U',N) + v^{\rm SIAM}_{\rm Hxc}(U_l - U', n_l),
\label{eq:vHxc_local}
\ee
where $U'$ is the {\em smallest} of all intersite interactions $U_{lm}$. Here we have defined the SIAM Hxc potential as 
\be
v^{\rm SIAM}_{\rm Hxc}(U, n) = \frac{U}{2}\left[1 - \frac{2}{\pi}\arctan\left(\frac{U(1-n)}{w_0 \gamma}\right)\right],
\label{vHxc_SIAM}
\ee
where $w_0$ is a numerical parameter. 
For most of our applications we choose the value $w_0=0.16$ which was 
found in Ref.~\cite{StefanucciKurth:15} to best reproduce results for the SIAM 
from reverse-engineering of rate equations. In contrast, 
for the calculation of the transmission spectral function of the double dot 
we choose $w_0=0.08$. 
 The CIM contribution to the Hxc potential is defined as 
\be
v^{\rm CIM}_{\rm Hxc}(U,N) = \sum_{k=0}^{2M-2} v^{\rm SIAM}_{\rm Hxc}(U, N-k) 
\ee
which depends only on the total occupation $N=\sum_l n_l$ of the MQD.

The structure of Eq.~(\ref{eq:vHxc_local}) reflects the physics of the model: the first term accounts for the charging energy associated with adding electrons to  a CIM of $M$ dots with interaction $U'$, while the second term captures the additional on-site correlation $U_l - U'$ specific to each dot. This decomposition is exact in the atomic limit and provides a natural generalization of the single-impurity functionals to the multi-dot case in parameter Regime I. For extensions to describe other regimes of the interaction parameters, see Ref.~\cite{SobrinoKurthJacob:20}.

The key ingredient for the i-DFT reconstruction of the interacting spectral function via Eq.~(\ref{tau_idft}) is the exchange-correlation bias $V_{\rm xc}$ in the STM limit. For a probe which connects only to a single site $l$, in parameter Regime I we found that the xc bias constructed from the Hxc potential as 
\bea
V_{\rm xc,l}(\{n_m\}, I) &=& v_{\rm Hxc,l}(\{n_m\}) - v_{\rm Hxc,l}(\{n_m + \delta_{lm} \tilde{I}/2\})\nonumber\\
&=& V_{\rm xc}^{\rm CIM}(U',N,\tilde{I}) + V_{\rm xc}^{\rm SIAM}(U_l-U',n_l,\tilde{I})
\label{eq:Vxc}
\eea
\noindent
 works well in the Coulomb blockade (CB) regime. In the first line of Eq.~(\ref{eq:Vxc}), 
$\delta_{lm}$ is the Kronecker delta and in the second line the terms have been collected according to Eq.~(\ref{eq:vHxc_local}). 
The construction of $V_{\rm xc,l}$ in terms of the corresponding $v_{\rm Hxc,l}$ reflects the fact that the probe current $\tilde{I}:=I/\gamma_{\rm P}$ through site $l$ shifts only the occupation at that site by $\tilde{I}/2$. 
The relationship between xc bias and Hxc potential according to Eq.~(\ref{eq:Vxc}) was found to hold for the SIAM in the CB regime and the STM 
limit \cite{JacobKurth:18}. In Eq.~(\ref{eq:Vxc}), the parameter $w_0$ entering is always fixed to $w_0=0.16$.

To access the Kondo regime, the different pieces of the xc bias are supplemented with a correction factor~\cite{KurthStefanucci:16,JacobKurth:18}, 
e.g., for the SIAM piece we write
\be
V_{\rm xc}^{{\rm SIAM, K}}(U, n_l, I) = \left[1 - a_K(U, I)\right] V_{\rm xc}^{\rm SIAM}(n_l, I) 
\label{Vxc_Kondo}
\ee
where
\be
a_K(U, I) = 1 - \frac{2}{\pi}\arctan\left[\left(\frac{U \tilde{I}}{4\gamma}\right)^{\!2} \frac{1}{w_1}\right] \;.
\label{eq:Kondo_pref}
\ee
In earlier work \cite{KurthStefanucci:16} we found that the value $w_1=0.16$ works 
well for the SIAM and therefore we use this value for the calculation of the 
local spectra. On the other hand, for the transmission spectral function of the 
DQD we use the value $w_1=0.48$. 
We note in passing that, following the ideas of 
Ref.~\cite{JacobStefanucciKurth:20,JacobKurth:26}, the parameter $w_1$ likely can be 
related to the quasi-particle weight, a known physical quantity of standard 
many-body theory.
The factor $a_K$ interpolates smoothly between $a_K \to 0$, where the Coulomb blockade xc bias is recovered, and $a_K \to 1$, where Kondo correlations are fully developed. For the CIM piece we proceed in an analogous way.

 In Sec.~\ref{ssec:transmission_spectra} we study the transmission spectral function of a double quantum dot in the Kondo regime but with a 
full off-diagonal coupling matrix ${\mathbf \Gamma}_{\rm P}$. In this geometry the probe couples to the double dot as a whole, rather than to a single site. Consequently, for each frequency one evaluates a single total probe current and introduces a single corresponding xc bias. For this coupling we need to adapt the xc bias. As a first guess (and keeping the structure of the 
xc bias as sum of different pieces of the interaction) we use
\be
V_{\rm xc}(\{n_m\},I) = V_{\rm xc}^{\rm CIM, K}(U_{12},N,\tilde{I}) + \frac{1}{2} 
\sum_{l=1}^2 V_{\rm xc}^{\rm SIAM, K}(U_l-U_{12},n_l,\tilde{I}) \;.
\label{eq:Vxc_transmission}
\ee
While this construction is approximate, we still expect this functional to be accurate in the vicinity of 
the particle-hole symmetric point, i.e., when $n_l \approx 1$ for $l \in \{1,2\}$, exactly the situation we will investigate in 
Sec.~\ref{ssec:transmission_spectra}.

With these ingredients, the self-consistent procedure proceeds in three steps: (i) the equilibrium densities $\{n_l\}$ are self-consistently determined from Eq.~(\ref{eq:n_digamma_stm}); (ii) for each frequency $\omega$, the current $I(V=\omega)$ is obtained from Eq.~(\ref{eq:Ired_digamma_stm}); and (iii) the interacting spectral function is reconstructed via Eq.~(\ref{tau_idft}). The nature of the current in step (ii) depends on the probe geometry. For a local probe coupled to site $j$ only, $({\mathbf \Gamma}_{\rm P})_{lm} = \gamma_{\rm P}\,\delta_{lj}\delta_{mj}$, and one solves for the local current $I(\omega)$ using the residues $q_k = \gamma_{\rm P}\,r_{k,jj}$. The quantity $\tau(\w)/\gamma_{\rm P}$ (see Eq.~(\ref{tau_idft})) then yields the local spectral function $A_{jj}(\omega)$. For the transmission spectral function, the probe couples to all sites through the full broadening matrix, $({\mathbf \Gamma}_{\rm P})_{lm} = \gamma_{\rm P}$ for all $l,m$, and one solves for the total current $I(\omega)$ using $q_k = \gamma_{\rm P}\sum_{lm}r_{k,lm}$. Via Eq.~(\ref{tau_idft}) we then obtain the transmission spectral function defined as
\begin{figure}
	\centering
	\includegraphics[width=\linewidth]{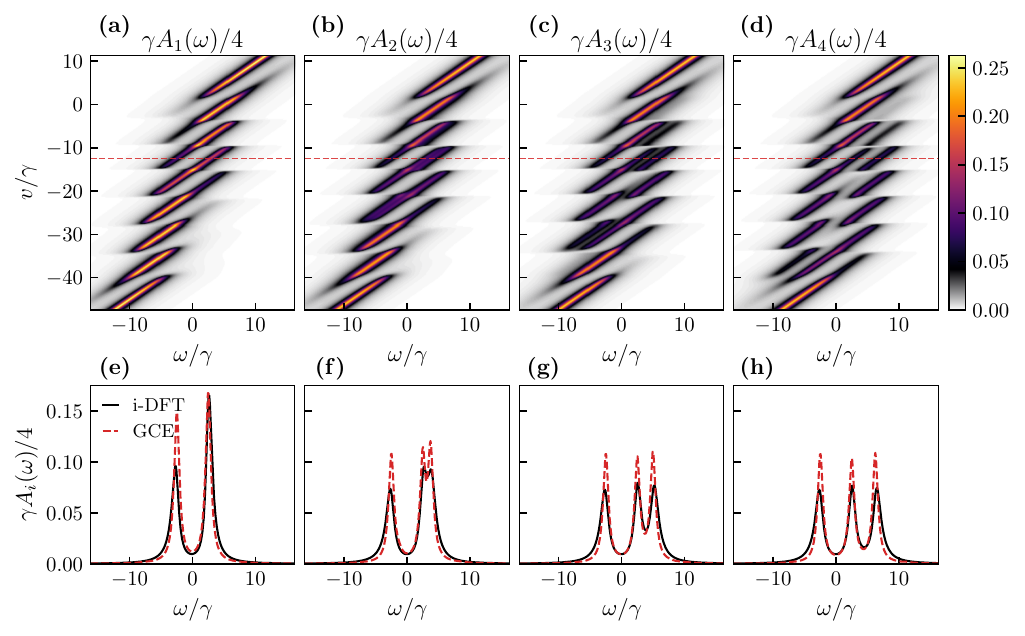}
	\caption{Local spectral functions of a quadruple quantum dot in the Coulomb blockade regime. Top row: color maps of the local spectral functions 
    $\gamma A_{ll}(\omega)/4$ as function of frequency $\omega$ and common gate voltage $v_l=v$. The dashed horizontal line indicates the gate voltage $v/\gamma=-12.5$ used for the line cuts below. Bottom row: comparison of $\gamma A_{ll}(\omega)/4$ obtained from i-DFT (solid black) and from GCE (dashed red) at the gate voltage indicated in the top row. For the spectral peaks of the GCE results  a Lorentzian broadening with parameter $\gamma$ was used. Parameters (in units of $\gamma$): $U_1=5$, $U_2=6.25$, $U_3=7.5$, $U_4=8.75$, $U_{lm}=5$, $T=0.5$, $t=0$. }
	\label{fig:figure1}
\end{figure}
\be 
T(\omega) = \frac{\gamma}{\gamma_{\rm P}} \tau(\w) \;.
\label{eq:transmission_specfunc}
\ee 

We emphasize that for any shape of the coupling matrices ${\mathbf \Gamma}$ and ${\mathbf \Gamma}_{\rm P}$, the pole 
re\-presentation of the KS Green function allows to reduce the integrals in the self-consistent equations to closed-form expressions, making the scheme  computationally highly efficient.

\subsection{Spectral Functions of Multiple Quantum Dots}
\label{ssec:mqd_spectra}

We begin by considering a system of $M=4$ quantum dots all of which are connected to the reservoir with equal and diagonal coupling without interdot hopping. We choose different on-site interactions $U_l$ and a common intersite 
interaction $U'=U_{lm}$ for all pairs $(l,m)$ which is not larger than any of the local interactions, 
$U'\leq U_l$. Figure~\ref{fig:figure1} shows the local spectral functions in the Coulomb blockade regime. The top row shows $\gamma A_{ll}(\omega)/4$ as a function 
of frequency and common gate voltage for each dot. As the gate is varied, electrons are added to the system, and the spectral weight rearranges accordingly. The 
different on-site interactions at each dot lead to distinct peak positions and spacings. The bottom row compares the i-DFT spectral functions at a fixed gate 
voltage with the exact grand canonical ensemble (GCE) result obtained via Lehmann representation where a Lorentzian broadening with parameter $\gamma$ has been 
applied to the spectral peaks.
 The GCE spectral functions, in particular the number, ordering,
  and the positions of the peaks,
  already correctly capture Coulomb blockade physics.  
 The i-DFT spectra show good agreement for these properties
  for 
all four dots, confirming the
proper construction of the xc functionals  of 
Sec.~\ref{ssec:mqd_spectra} for the multi-dot case.
 However, there are still residual differences in peak shapes and
  spectral weights. 
As mentioned above, the xc bias (\ref{eq:Vxc}) employed here is constructed for $U_{lm} < U_j$ (referred to as 
Regime~I in Ref.~\cite{SobrinoKurthJacob:20}). For other parameter 
regimes, which are arguably less physically relevant, additional steps may develop in the $V_{xc,l}$  in the (hyper) plane of the occupation vector $\{n_l\}$ 
and corresponding probe current $I$. This situation requires more elaborate functionals and is beyond the scope of this work.

\begin{figure}
	\centering
	\includegraphics[width=\linewidth]{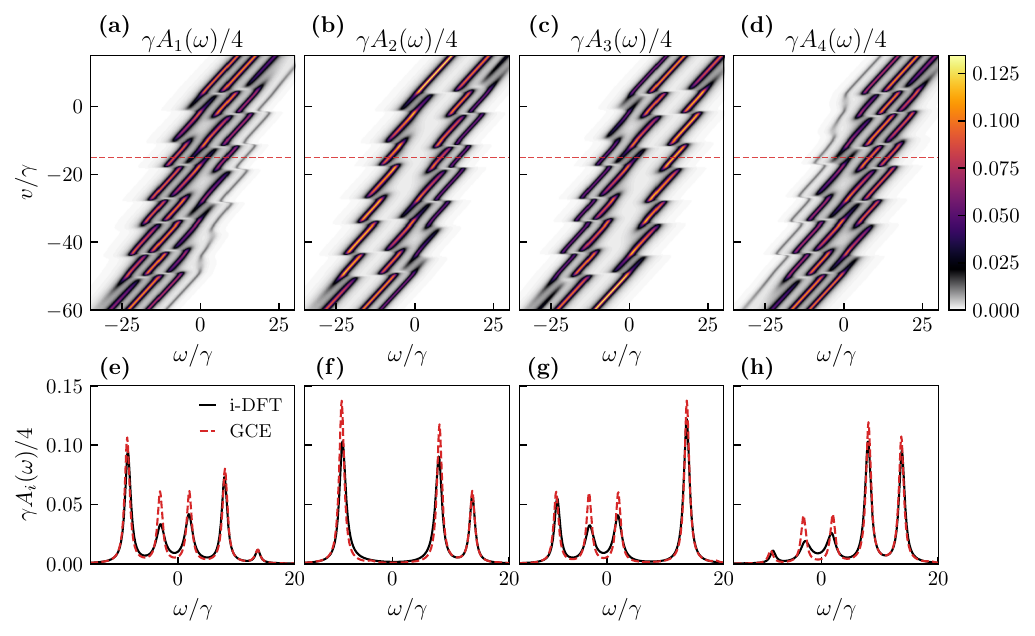}
	\caption{Local spectral functions of a quadruple quantum dot system with interdot hopping and CIM interaction $U_l=U_{lm}=U$ in the Coulomb blockade regime. Top row: colour-map of $\gamma A_{ll}(\omega)/4$ as a function of frequency $\omega/\gamma$ and average gate voltage $v/\gamma$ for each dot $l=1,\ldots,4$. The individual gate voltages are $v_l = v + \delta_l$ with shifts $\delta_l/\gamma = 2l$. The dashed horizontal line indicates the gate voltage $v/\gamma=-15$ used for the line cuts below. Bottom row: comparison of $ A_{ll}(\omega)/4$ obtained from i-DFT (solid black) and GCE (dashed red) at the gate voltage indicated in the top row. 
    For the spectral peaks of the GCE results  a Lorentzian broadening with parameter $\gamma$ was used. Parameters (in units of $\gamma$): $U=5$, $t=5$, $T=0.5$.}
	\label{fig:figure2}
\end{figure}

Next we turn on nearest-neighbour interdot hopping ($t_{l,l+1} = t$) but choose all on-site and intersite interactions equal (CIM). Figure~\ref{fig:figure2} shows results for $M=4$ dots all of which are connected to the reservoir with equal and diagonal coupling. We use $t/\gamma = 5$, 
temperature $T/\gamma=0.5$, and a gate detuning  $v_l = v + \delta_l$ which breaks the symmetry between the dots. The hopping hybridizes 
the spectral features across dots, leading to richer peak structures compared 
to the decoupled case: peaks that were degenerate in the absence of hopping now split, and spectral weight is redistributed among the dots. Despite this additional complexity, the i-DFT results 
reproduce the positions of the GCE peaks as well as the qualitative redistribution of spectral weight among the dots, while residual differences in peak shapes and weights similar to those of Fig.~\ref{fig:figure1} remain. This shows that the framework handles interdot hopping in the constant interaction case without modification of the xc functionals.
\begin{figure}
	\centering
	\includegraphics[width=\linewidth]{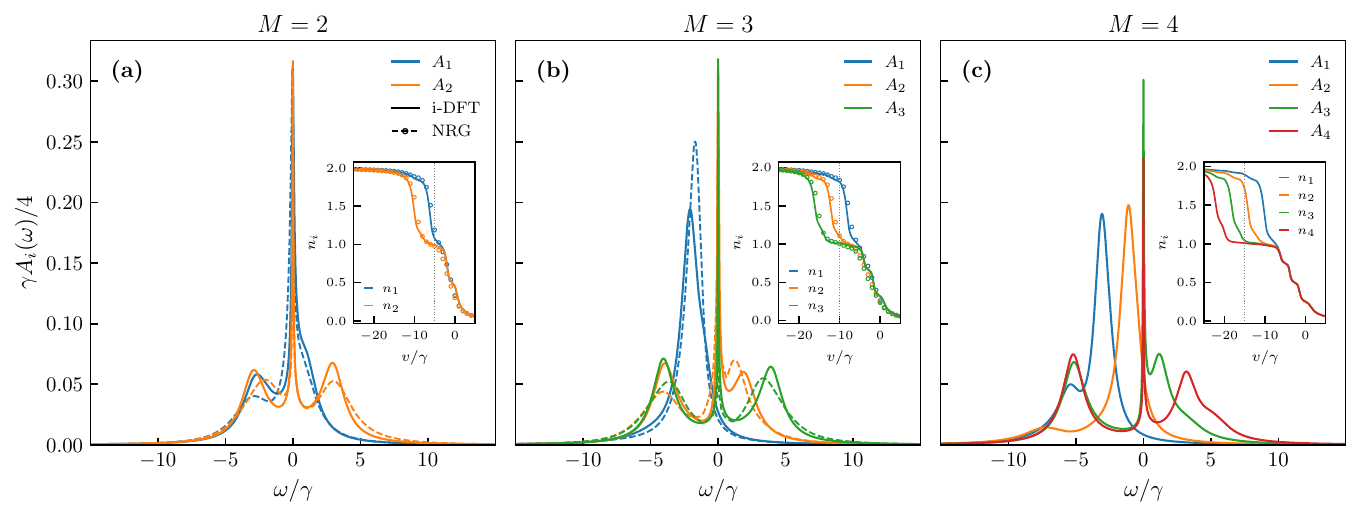}
	\caption{Local spectral functions of multiple quantum dots without interdot hopping in the Kondo regime. (a) $M=2$, (b) $M=3$, and (c) $M=4$: $\gamma A_l(\omega)/4$ for each dot $l$ obtained from i-DFT (solid lines). For $M=2$ and $M=3$ the dashed lines show the corresponding NRG spectral functions \cite{NRG_calc}. A Kondo peak is visible near $\omega=0$. Insets: site occupations $n_l$ as a function of the common gate voltage $v/\gamma$; the vertical dashed line indicates the gate voltage at which the spectral functions are evaluated. Parameters (in units of $\gamma$): $U_{lm}=2$, $T=10^{-5}$; (a) $U_1=4$, $U_2=6$, $v=-5$; (b) $U_1=4$, $U_2=6$, $U_3=8$, $v=-10$; (c) $U_1=4$, $U_2=6$, $U_3=8$, $U_4=10$, $v=-15$.}
	\label{fig:figure3}
\end{figure}

Finally, we explore the Kondo regime by lowering the temperature and reducing the interaction strengths. From an i-DFT perspective, this is achieved by including the Kondo prefactor of Eq.~(\ref{eq:Kondo_pref}), which has already been shown to correctly capture the Kondo peak in the low-temperature limit for the SIAM~\cite{KurthStefanucci:16}. An extension of this construction which captures the evolution of the peak as a function of temperature has been 
developed~\cite{KurthStefanucci:16}, but in order to keep a simple structure we restrict ourselves here to the low-temperature 
limit, such that the results below do not address the universal scaling with $T/
T_K$ where $T_K$ is the Kondo temperature. Figure~\ref{fig:figure3} shows local spectral functions for MQDs with $M=2$, $3$, and $4$ dots without interdot 
hopping, at gate voltages chosen such that some dots are near half-filling (see insets). In addition to the Coulomb blockade 
side peaks, a narrow Kondo resonance is clearly visible near $\omega = 0$ for the dots around half-filling.

Figure~\ref{fig:figure3} shows local spectral functions for MQDs with $M=2$, $3$, and $4$ dots without interdot hopping, at gate voltages chosen such that some dots
are near half-filling (see insets). For $M=2$ and $M=3$ the i-DFT results are compared with NRG calculations~\cite{NRG_calc}. In addition to the Coulomb
blockade side peaks, a narrow Kondo resonance is visible near $\omega=0$ for the dots around half-filling. i-DFT reproduces qualitatively the positions of the side 
peaks, the Kondo resonances and the gate dependence of the occupations. Here we emphasize again that, compared to NRG the i-DFT results were obtained 
at negligible computational cost (see also discussion at the end of Sec.~\ref{ssec:idft_stm}). For instance, the NRG calculations for the M=3 MQD required
computational resources which were five orders of magnitude larger than
the ones for the corresponding i-DFT calculations.

\subsection{Transmission Spectral Functions of Double Dots}
\label{ssec:transmission_spectra}
We now turn to a qualitatively different application of the STM framework: the computation of transmission spectral functions, see 
definition of Eq.~(\ref{eq:transmission_specfunc}). As discussed in Sec.~\ref{ssec:idft_stm}, the flexibility of the probe coupling matrix ${\mathbf \Gamma}_{\rm P}$ allows one to project out not only local but also non-local spectral information. In particular, for a double quantum dot ($M=2$) connected to a reservoir with off-diagonal broadening $\Gamma_{12} \neq 0$ (corresponding to a lead coupling to both dots), the transmission spectral function $T(\omega)$ is accessed by choosing the probe coupling proportional to the reservoir broadening matrix.

We consider a double quantum dot at particle-hole symmetry with equal on-site interactions $U_1 = U_2 = U$, intersite interaction $U_{12} = U - \Delta U$, and a detuning $\Delta\varepsilon = v_2 - v_1$ between the two dot levels. This system has been studied in Ref.~\cite{kleeorin2018quantum} using the numerical renormalization group (NRG), where a quantum phase transition (QPT) was identified as a function of the ratio $\Delta\varepsilon/\Delta U$ which plays the role af a magnetic interaction: for $\Delta\varepsilon/\Delta U < 1$ the interaction is ferromagnetic-like (FM) and the system exhibits a broad Kondo resonance, while 
for $\Delta\varepsilon/\Delta U > 1$ the interaction is antiferromagnetic 
(AFM) and the developing phase is characterized by a suppression of the transmission spectral function at zero frequency.

Within i-DFT, the KS transmission spectral function $T_s(\w)$ (defined as in to Eq.~(\ref{eq:transmission_specfunc}) but with 
the KS spectral function matrix ${\mathbf A}_s$) for the double dot with onsite potentials chosen as 
$v_1=v-\Delta\varepsilon/2$ and $v_2=v+ \Delta\varepsilon/2$ with $v=-U/2 - U_{12}$ can be analyzed analytically. In this case, by symmetry the KS on-site energies satisfy $v_{s,2} = -v_{s,1}$. Using the spectral representation of the KS Green function derived in the Appendix \ref{sec:appendix1}, the KS transmission spectral function can then be written as
\be
T_s(\omega) = \frac{2\gamma^2}{\pi}\,\frac{\omega^2}{(\omega^2 - v_{s,1}^2)^2 + \gamma^2\omega^2}\;.
\label{eq:Ts_analytic}
\ee
From this expression one sees that the KS transmission exactly vanishes at $\omega = 0$ whenever $v_{s,1} \neq 0$, i.e., whenever the two KS levels are split. On the other hand, if $v_{s,1}=0$, i.e., at particle-hole symmetry, then $T_s(\w=0)= 2/\pi$. 
The maxima occur at $\omega = \pm |v_{s,1}|$, where $T_s = 2/\pi$. A natural definition of the width of the central dip is obtained from the full width at half maximum of the central antiresonance which yields
\be
\gamma_{\rm dip} = \sqrt{\gamma^2 + 4v_{s,1}^2} - \gamma\;.
\label{eq:dip_width}
\ee
The visibility of the dip in the KS transmission is thus controlled by the competition between the KS level splitting $v_{s,1}$ and the broadening $\gamma$: for small $|v_{s,1}|$ the dip is extremely narrow, while increasing $|v_{s,1}|$ makes it progressively resolvable.

\begin{figure}
	\centering
    \includegraphics[width=\linewidth]{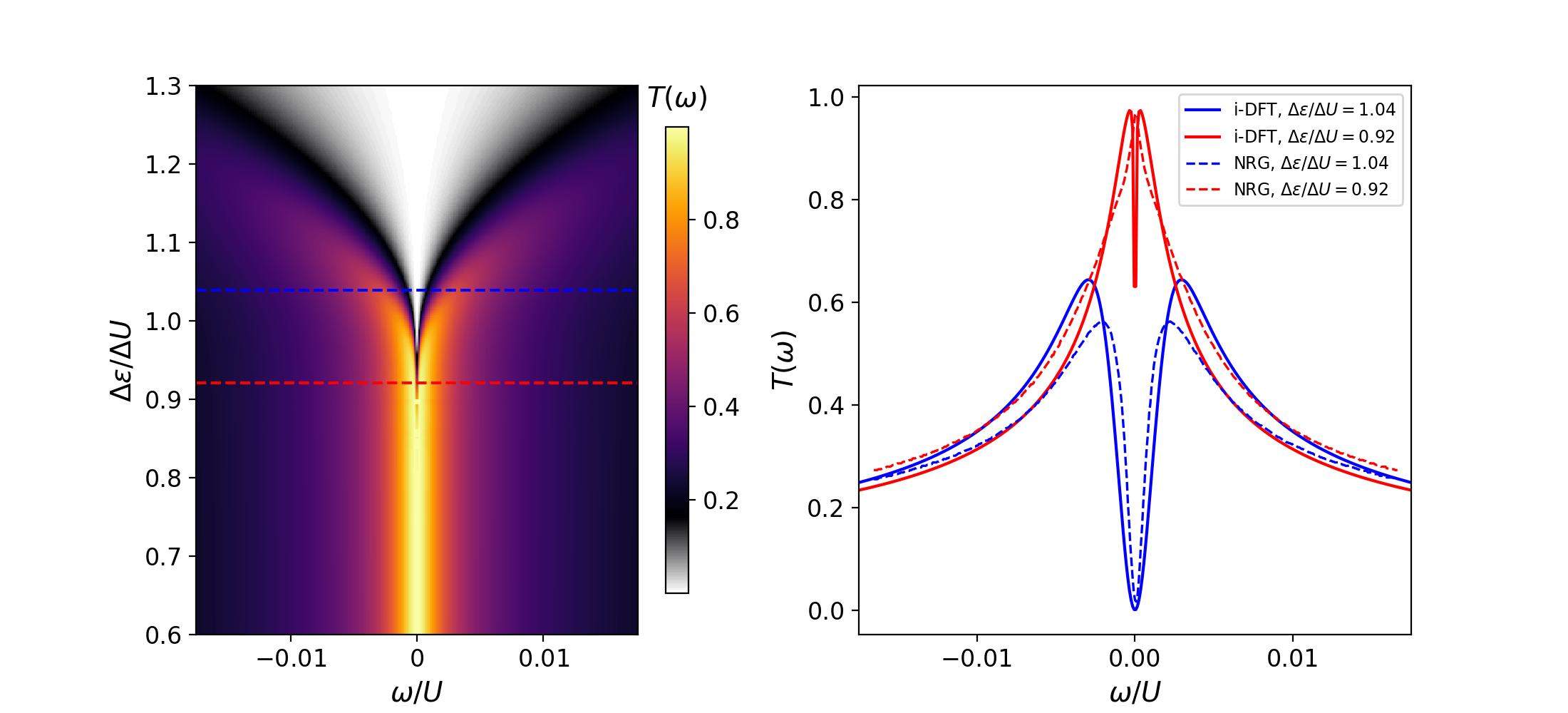}
	\caption{
    Transmission spectral function $T(\omega)$ of a double quantum dot as a function of frequency $\omega$ and level detuning $\Delta\varepsilon/\Delta U$, computed within i-DFT. Left: full (off-diagonal) coupling matrix, showing the change from a broad Kondo resonance ($\Delta\varepsilon/\Delta U < 1$) to a suppressed zero-frequency transmission ($\Delta\varepsilon/\Delta U > 1$). Right: line cuts of \(T(\omega)\) for \(\Delta\varepsilon/\Delta U=0.92\) and
\(\Delta\varepsilon/\Delta U=1.04\), compared with the NRG results of
Ref.~\cite{kleeorin2018quantum}. Parameters: $U/\gamma = 15$, $\Delta U = U/3$, $T = 3\cdot10^{-6}\,U$.}
	\label{fig:transmission}
\end{figure}

From Eq.~(\ref{tau_idft}) we can derive the interacting transmission spectral function as
\be
T(\w) = \frac{T_s(\w+ V_{\rm xc})}{1- \frac{\partial V_{\rm xc}}{\partial I}\frac{\gamma_{\rm P}}{\gamma} T_s(\w + V_{\rm xc})} \;.
\label{eq:transmission_int}
\ee
Of course, at zero bias ($V=\w=0$) both the current as well as the xc bias vanish. Therefore, for the situation studied above 
($v_1=v-\Delta\varepsilon/2$, $v_2=v+\Delta\varepsilon/2$), by Eq.~(\ref{eq:transmission_int}), at $\w=0$  not only the KS but also the interacting spectral function vanishes for $\Delta\varepsilon \neq 0$. On the other hand, at particle-hole symmetry 
($\Delta\varepsilon=0$) both quantities remain finite.  In fact, if the derivative of $V_{\rm xc}$ with respect to $I$ vanishes at $I=0$ (as is the case in the approximation we use), at particle-hole symmetry the KS and interaction transmission spectral function take the {\em same} non-zero value.

The left panel of Figure~\ref{fig:transmission} shows the colormap of the i-DFT transmission $T(\omega)$ as a function of frequency and $\Delta\varepsilon/\Delta U$ for a double quantum dot in the strongly correlated regime.  The i-DFT transmission reproduces the change of the transmission line shape across the transition reported in Ref.~\cite{kleeorin2018quantum}: for $\Delta\varepsilon/\Delta U < 1$, a broad Kondo resonance dominates the zero-frequency transmission, while for $\Delta\varepsilon/\Delta U > 1$ the transmission is strongly suppressed near $\omega = 0$. Around the critical region $\Delta\varepsilon/\Delta U \sim 1$, a sharp dip develops in the Kondo resonance. In the i-DFT picture, this transition is driven by the self-consistent KS level splitting: as $\Delta\varepsilon/\Delta U$ crosses around a value close to unity, the KS levels split apart and the antiresonance described by Eq.~(\ref{eq:Ts_analytic}) opens up. The scale $\gamma_{\rm dip}$ of Eq.~(\ref{eq:dip_width}) characterizes the
width of the central interference antiresonance and decreases as 
$\Delta\varepsilon/\Delta U$ approaches unity from above. 

The right panel shows line cuts of the colormap at fixed values of $\Delta\varepsilon/\Delta U$, together with the corresponding NRG results of Ref.~\cite{kleeorin2018quantum}, showing quantitative
agreement between i-DFT and NRG for representative values on the two sides of the transition. For $\Delta\varepsilon/\Delta U=0.92$, both methods display a Kondo resonance centered at the Fermi level, whereas for $\Delta\varepsilon/\Delta U=1.04$ both show a pronounced suppression of $T(\omega)$ at $\omega=0$ inside a broader resonance. The i-DFT results also resolve the formation of the central dip already slightly below, around $\Delta\varepsilon/\Delta U\simeq 0.9$.

\section{Conclusions}
\label{sec:conclusions}

We have presented an extension of steady-state density functional theory
(i-DFT) for the computation of spectral and transmission properties of
interacting multiple quantum dots coupled to electronic reservoirs. The
approach builds on the ideal STM limit of i-DFT, where the differential
conductance through an infinitesimally weakly coupled probe gives direct access to
equilibrium spectral information. By allowing for general probe-coupling
matrices, the framework provides access not only to local spectral functions
but also to non-local components of the many-body spectral function matrix,
which are essential for describing interference effects and for reconstructing
transmission spectral functions.

The central practical result is a reconstruction formula which expresses the
interacting spectral quantity in terms of the corresponding Kohn-Sham quantity as well as the 
the exchange-correlation bias $V_{\rm xc}$. In the STM limit, the density
problem reduces to an equilibrium DFT calculation, while the current equation
decouples and can be solved independently at each bias/frequency. The resulting
numerical scheme is therefore highly efficient: after solving $M$ coupled
equations for the equilibrium densities, each spectral quantity 
requires only the solution of an independent scalar nonlinear equation (for each frequency). 
This represents a substantial reduction in computational cost compared to many-body approaches.

The formal framework is independent of the specific form of the interaction in
the central region. Its practical implementation, however, requires
approximations for the Hxc potentials and xc bias functionals, whose form
depends on the interaction under consideration. In this work, we have
constructed such approximate functionals for multi-dot systems with
density-density interactions. We applied these functionals to several
representative multiple-dot geometries. In the Coulomb blockade regime, the
resulting local spectral functions of quadruple quantum dots  correctly reproduce the number,
and positions of the peaks of the many-body reference. This agreement holds both for dots
 with both vanishing and finite interdot hopping in the
constant-interaction limit, showing that the present functionals capture the splitting and
redistribution of spectral weight induced by the interdot hybridization between dots.

We have also applied the method to multiple quantum dots in the Kondo regime.
For systems with $M=2,3,$ and $4$ dots, the i-DFT spectra display narrow
resonances at the Fermi level whenever the corresponding dots are close to
half filling, together with the expected Coulomb-blockade side peaks. 
For $M=2$ and $M=3$, comparison with NRG shows that i-DFT qualitatively reproduces
the positions of the side peaks, the Kondo resonances and the gate dependence of
the occupations, thus demonstrating its usefulness as a low-cost route to strongly correlated
spectral properties in systems beyond the single-impurity limit.

Finally, we have computed the transmission spectral function of a strongly
correlated double quantum dot with an off-diagonal reservoir broadening matrix. The i-DFT results reproduce 
representative NRG line shapes on the two sides of the transition previously identified by NRG: a broad Kondo resonance for $\Delta\varepsilon/\Delta U<1$, and a
strong suppression of the zero-frequency transmission for
$\Delta\varepsilon/\Delta U>1$. Within our formulation, this
suppression can be understood analytically as arising from a splitting of the effective dot levels, which opens an interference
antiresonance in the transmission. The corresponding dip width provides a spectral measure of the emergent low-energy scale associated with the antiresonance.

Overall, the results demonstrate that i-DFT in the ideal STM limit provides a
compact, flexible, and computationally efficient route to spectral and
transmission properties of interacting nanoscale systems. As in any
density-functional approach, the quantitative accuracy ultimately depends on
the quality of the approximate functionals. Nevertheless, the
benchmarks and applications presented here show that physically motivated
functionals already capture Coulomb blockade, finite interdot hopping, Kondo
resonances, and interference-induced splitting of this resonnce in
multiple quantum dots. Natural future directions include 
finite-temperature functionals able to describe universal $T/T_K$ scaling,
functionals for underscreened fixed points for which the xc bias must exhibit a
non-analytic low-current dependence in order to reproduce singular-Fermi-liquid
cusps, interdot charge-transfer contributions to the Hxc potential 
beyond Regime I, and nonempirical prescriptions for the crossover scales $w_0$
and $w_1$.

\section*{Data Availability}
The data that support the findings of this article are openly available in Ref.~\cite{SobrinoKurth_ZenodoData_iDFT_MQD_2026}.

\section*{Acknowledgements}

N.S. acknowledges the European Union under
the Horizon Europe research and innovation programme (Marie Skłodowska-Curie grant agreement no. 101148213, EATTS). 
S.K. acknowledges financial support from the Basque Government through the Elkartek program (project CICe2025, grant number KK2025-00054). 
We acknowledge technical support provided by SGIker (Scientific Computing Services
UPV/EHU).

\begin{appendix}
\section{Bi-orthogonal spectral representation of the resolvent}
\label{sec:appendix1}

We briefly recall the spectral representation of the resolvent of a
diagonalizable non-Hermitian operator \(\hat H\), a standard tool in the wide-band-limit NEGF description of embedded non-interacting junctions~\cite{tuovinen2014time,ridley2022many}.
Let \(w_k\) be its eigenvalues and let \(|r_k\rangle\) and \(|l_k\rangle\)
denote the corresponding right and left eigenvectors,
\begin{equation}
\hat H|r_k\rangle = w_k |r_k\rangle,
\qquad
\hat H^\dagger |l_k\rangle = w_k^* |l_k\rangle ,
\end{equation}
or equivalently \(\langle l_k|H = w_k \langle l_k|\).
For a diagonalizable matrix, the right and left eigenvectors may be chosen
bi-orthogonal,
\begin{equation}
\langle l_k|r_q\rangle = \delta_{kq}\langle l_k|r_k\rangle .
\end{equation}
The corresponding resolution of the identity is then
\begin{equation}
\hat{I}
=
\sum_k
\frac{|r_k\rangle\langle l_k|}
{\langle l_k|r_k\rangle}.
\end{equation}
Indeed, acting on any right eigenvector \(|r_q\rangle\), the right-hand side
returns \(|r_q\rangle\), and the set \(\{|r_q\rangle\}\) is complete by
assumption.
Using the same projectors, the matrix itself can be written as
\begin{equation}
\hat H
=
\sum_k
w_k
\frac{|r_k\rangle\langle l_k|}
{\langle l_k|r_k\rangle}.
\end{equation}
It follows immediately that any function of \(H\) which is regular on the
spectrum is represented as
\begin{equation}
f(\hat H)
=
\sum_k
f(w_k)
\frac{|r_k\rangle\langle l_k|}
{\langle l_k|r_k\rangle}.
\end{equation}
Choosing \(f(z)=(\omega-z)^{-1}\), with \(\omega\) away from the spectrum of
\(\hat H\), gives the resolvent
\begin{equation}
(\omega\hat{I}-\hat H)^{-1}
=
\sum_k
\frac{|r_k\rangle\langle l_k|}
{(\omega-w_k)\langle l_k|r_k\rangle}.
\label{eq:biorth_resolvent}
\end{equation}
Taking matrix elements in the site basis therefore yields
\begin{equation}\label{eq:app_Gij}
	\big[(\omega\mathbb{I}-\mathbf H)^{-1}\big]_{lm}
	=\sum_k\frac{(\mathbf{r}_k)_l\,(\mathbf{l}_k)_m^*}{(\omega-w_k)\,\mathbf{l}_k^\dagger\,\mathbf{r}_k},
\end{equation}
where $(\mathbf{r}_k)_l$ and $(\mathbf{l}_k)_m^*$ denote the $l$th and $m$th components of the right and conjugated left eigenvectors, respectively.

For $M=2$ (two-site case), we consider the general Hamiltonian
\begin{equation}
	\mathbf H_{\mathrm{eff}}=
	\begin{pmatrix}
		v_{s,1} & t\\
		t & v_{s,2}
	\end{pmatrix}
	-\frac{i}{2}
	\begin{pmatrix}
		\Gamma_{11} & \Gamma_{12}\\
		\Gamma_{21} & \Gamma_{22}
	\end{pmatrix}.
\end{equation}
Defining the auxiliary quantities
\begin{equation}
	\bar{v} = \frac{v_{s,1}+v_{s,2}}{2}-\frac{i}{2}\frac{\Gamma_{11}+\Gamma_{22}}{2}, \qquad
	\Delta = \frac{v_{s,1}-v_{s,2}}{2}-\frac{i}{2}\frac{\Gamma_{11}-\Gamma_{22}}{2}, \qquad
	\tilde{t} = t - \frac{i}{2}\Gamma_{12},
\end{equation}
the two eigenvalues of $\mathbf H_{\mathrm{eff}}$ are
\begin{equation}
	w_\pm=\bar{v}
	\pm \sqrt{\Delta^2+\tilde{t}^2}.
\end{equation}
Each matrix element of the resolvent admits the partial-fraction decomposition
\begin{equation}
	\big[(\omega\mathbb{I}-\mathbf H_{\mathrm{eff}})^{-1}\big]_{ij}=\frac{r^{(+)}_{ij}}{\omega-w_+}+\frac{r^{(-)}_{ij}}{\omega-w_-},
\end{equation}
with residues
\begin{align}
	r^{(\pm)}_{11}&=\frac{w_\pm-\big(v_{s,2}-i\Gamma_{22}/2\big)}{w_\pm-w_\mp},
	&
	r^{(\pm)}_{22}&=\frac{w_\pm-\big(v_{s,1}-i\Gamma_{11}/2\big)}{w_\pm-w_\mp},\nonumber\\
	r^{(\pm)}_{12}&=\frac{\tilde{t}}{w_\pm-w_\mp},
	&
	r^{(\pm)}_{21}&=\frac{\tilde{t}}{w_\pm-w_\mp}.
\end{align}

\end{appendix}

\bibliographystyle{SciPost_bibstyle} 

\nolinenumbers

\end{document}